\documentclass[10pt,a4paper]{article}
\usepackage[margin={20mm,28mm}]{geometry}
\usepackage{amsmath,amssymb,mathrsfs,cite}
\usepackage{color}
\usepackage{graphicx}
\usepackage[absolute,overlay]{textpos}
\usepackage{ascmac}
\usepackage{hyperref}

\allowdisplaybreaks
\numberwithin{equation}{section}

\def\no{\notag\\}
\def\hana{\mathscr}
\newcommand{\tr}{\operatorname{tr}}

\begin{document}
\begin{titlepage}
\begin{flushright}
UTHEP-805
\end{flushright}
\vspace{12mm}
\begin{center}
{\LARGE \bf
Classical BPS M5-brane
on the plane wave background
}
\end{center}
\vspace{7mm}
\begin{center}
Yuhma Asano$^{a,b,}$\footnote{
asano@het.ph.tsukuba.ac.jp
},
Goro Ishiki$^{a,b,}$\footnote{
ishiki@het.ph.tsukuba.ac.jp
},
Yuto Yoshida$^{a,}$\footnote{
yoshida@het.ph.tsukuba.ac.jp
}

\par \vspace{7mm}
$^a${\it
Graduate School of Science and Technology, University of Tsukuba,\\
1-1-1 Tennodai, Tsukuba, Ibaraki 305-8571, Japan
}\\

$^b${\it
Tomonaga Center for the History of the Universe, University of Tsukuba,\\
1-1-1 Tennodai, Tsukuba, Ibaraki 305-8571, Japan
}\\

\end{center}
\vspace{15mm}
\begin{abstract}\noindent
We consider the bosonic theory for a single M5-brane on the plane-wave background
and derive a family of BPS solutions 
with non-zero components of the angular momentum.
By explicit construction of a BPS solution,
we find the solution describes 
an ellipsoidal five-brane rotating without changing its shape.
The methodology developed in this paper is expected to 
provide a strategy for obtaining 
the BPS solutions that correspond to a BPS sector in the dual gauge theory,
such as the BMN matrix model.

\end{abstract}
\setcounter{footnote}{0}
\end{titlepage}

\section{Introduction}

The M5-brane is still a mysterious object 
whose theory in the Lagrangian formalism is not known 
for multiple branes---even the existence of such a formalism is questionable.

While the direct approach to study a system of multiple M5-branes is limited,
there is another approach via Matrix theory \cite{Banks:1996vh}.
Matrix theory was conjectured as a non-perturbative formulation of M-theory,
in which a theory on M-branes is realized as 
quantum theory around a classical solution.
In particular, 
it is conjectured that
the mass-deformed matrix theory with maximal supersymmetry,
known as the BMN matrix model (plane-wave matrix model) \cite{Berenstein:2002jq}, 
realizes transverse M5-branes as a vacuum of the matrix model \cite{Maldacena:2002rb}.
In fact, it was confirmed that 
the eigenvalue distribution of a quarter-BPS operator
in the matrix model 
reproduces the spherical geometry of multiple stacks of transverse M5-branes
on the plane-wave background, 
including the correct values of their radii,
in the decoupling limit \cite{Asano:2017xiy,Asano:2017nxw}.

In the matrix model, 
the expectation value of any function of the quarter-BPS operator 
can be computed by a simple matrix integral, 
as a result of the supersymmetric localization method \cite{Asano:2012zt}.
Then, the eigenvalue distribution of the BPS operator 
is determined as a saddle point of
the effective action associated with the matrix integral.
Since the effective action characterizes the quarter-BPS sector,
fluctuation of the eigenvalues around the saddle point corresponds to
quantum fluctuation in the quarter-BPS sector of the original matrix model.

On the other hand,
in the M-theoretic picture,
the geometry of the transverse M5-branes is obtained as a classical solution 
of the theory on a single M5-brane,
which is known to have a covariant action with $\kappa$ symmetry \cite{Pasti:1997gx,Bandos:1997ui} with the addition of an auxiliary scalar field.
In particular, the spherical geometry with zero light-cone energy 
was obtained as a classical solution of the bosonic part of the five-brane theory
without the chiral two-form field and the auxiliary field
\cite{Maldacena:2002rb,Asano:2017xiy,Asano:2017nxw}.
Since the eigenvalue distribution at the saddle point
in the matrix model
reproduces the spherical geometry,
we expect the fluctuation in the BPS sector of the matrix model
should correspond to the deformed geometry described by a BPS solution.
In the literature,
there are some BPS solutions\footnote{
See Ref.~\cite{Ho:2012dn} for BPS solutions in a different formalism \cite{Ho:2008nn,Ho:2008ve}.
} obtained for a single M5-brane on a background \cite{Howe:1997ue,Howe:1997hx,Lust:1999pq,Michishita:2000hu,Youm:2000kr},
but there is none found for the plane-wave background.

In this paper,
we develop a methodology for 
obtaining BPS solutions 
in the five-brane theory on the plane-wave background,
towards understanding of
their precise correspondence to the quarter-BPS sector of the matrix model.
The paper is organized as follows.
In section 2, we review how the spherical geometry is derived 
as a classical solution with zero light-cone energy.
Then, in section 3, we compute the BPS saturation condition
and obtain BPS solutions labeled by four components of the angular momentum.
Finally, we summarize the results in section 4.

\section{Spherical solution on the plane wave background}
Let us first review the classical solution to the M5-brane theory 
with zero light-cone energy
on the plane-wave background,
\begin{align}
ds^2 &=g_{\mu\nu}dx^\mu dx^\nu
=-2dx^+dx^-+\sum_{A=1}^9 dx^Adx^A-\left(\frac{\mu^2}{9}\sum_{i=1}^3x^ix^i+\frac{\mu^2}{36}\sum_{a=4}^9x^ax^a\right)dx^+dx^+ ,
\nonumber \\
F_4&=\mu\, dx^1\wedge dx^2\wedge dx^3\wedge dx^+ ,
\label{planewave-metric}
\end{align}
where $\mu$ is a constant flux parameter, which we assume non-zero in the paper.
We start with the bosonic five-brane action on this background
\begin{align}
S =-T_{\text{M5}} \int d^6\sigma\left(-h\right)^{1/2}+T_{\text{M5}}\int C_6,
\end{align}
where $dC_6=\ast F_4$ and
we set the chiral two-form field, the auxiliary field and the fermionic fields 
to zero.
Here, $h$ is the determinant of the induced metric
of the embedding $X^\mu(\sigma)$,
\begin{align}
 h_{mn}=g_{\mu\nu}
 \partial_mX^\mu\partial_nX^\nu
 ,
\end{align}
where 
$\sigma^m$ are the world-volume coordinates with $m,n=0,\cdots,5$
and the target space indices $\mu,\nu$ run over the values $+,-,1,\cdots,9$.
The overall constant $T_{\text{M5}}$ is the tension of the M5-brane, 
\begin{align}
T_{\text{M5}}= \left(2\pi\right)^{-5}l_{\text{p}}^{-6},  
\end{align}
where $l_{\text{p}}$ represents the Plank length. 

Just like standard perturbative string theory,
we rewrite the Nambu-Goto-type action to the Polyakov type. 
We first note that the constraints are
\begin{align}
0&=P^\mu \partial_\alpha X_\mu, \label{kousoku1}\\
0&=P^2+T^2_{\text{M5}}\det h_{\alpha\beta}=P^2+\frac{T^2_{\text{M5}}}{5!}\left\{X^{A_1},\cdots,X^{A_5}\right\}^2, \label{kousoku2}
\end{align}
in the Hamiltonian system
by a Legendre transformation of the Nambu-Goto-type action, 
where $\alpha,\beta=1,\cdots,5$.
Then,
we obtain the Polyakov-type action 
by another Legendre transformation 
with the constraints taken into account,
by integrating out $X^-$,
as
\begin{align}
S&=\frac{1}{4\lambda}\int d^6\sigma\left(h_{00}-2\lambda^\alpha h_{\alpha 0}+\lambda^\alpha\lambda^\beta h_{\alpha\beta}-4\lambda^2 T_{\text{M5}}^2\det h_{\alpha\beta}\right)+T_{\text{M5}}\int C_6\no
&=\frac{p^+}{2V_5}\int d^6\sigma \left(D_0X^AD_0X_A-\frac{T^2_{\text{M5}}}{5!}\left(\frac{V_5}{p^+}\right)^2\left\{X^{A_1},\cdots,X^{A_5}\right\}^2-\left(\frac{\mu^2}{9}X^iX_i+\frac{\mu^2}{36}X^aX_a\right)\right)\no
&\qquad+\frac{\mu T_{\text{M5}}}{6!}\int d^6\sigma\epsilon_{a_1\cdots a_6}X^{a_1}\left\{X^{a_2}\cdots X^{a_6}\right\},
\label{P-action}
\end{align}
where $\epsilon_{a_1\cdots a_6}$ is the totally antisymmetric tensor with $\epsilon_{456789}=1$,
$D_0$ is the covariant derivative associated with $\lambda^\alpha$, and 
$\lambda^\alpha$ and $\lambda$ are the Lagrange multipliers for the constraint (\ref{kousoku1})
and 
(\ref{kousoku2}), 
respectively.
Here, the 5-bracket is defined by 
\begin{align}
\left\{X^{A_2},\cdots ,X^{A_6}\right\} = \epsilon^{\alpha_1\cdots \alpha_5} \partial_{\alpha_1}X^{A_2}\cdots \partial_{\alpha_5}X^{A_6},
\end{align}
where $\epsilon^{\alpha_1\cdots\alpha_5}$ is the totally antisymmetric tensor for the spatial part of the world-volume, with $\epsilon^{12345}=1$.
To derive the final expression of \eqref{P-action}, 
we chose the light-cone gauge $X^+=\sigma^0$
and fix $\lambda$ to a constant,
which turns out to be related to the total light-cone momentum $p^+$ by 
\begin{align}
p^+=\int d^5\sigma\,P^+=\frac{V_5}{2\lambda},
\end{align}
where $V_5$ is the volume of a unit 5-sphere. 

The light-cone Hamiltonian is given by
\begin{align}
H&= \frac{V_5}{2p^+}\int d^5\sigma \left(P^AP_A+\frac{T^2_{\text{M5}}}{5!}\sum_{(A_1,\cdots,A_5)\neq(a_1,\cdots,a_5)}\left\{X^{A_1},\cdots,X^{A_5}\right\}^2+\left(\frac{\mu p^+}{3V_5}\right)^2X^iX_i\right.\no
& \left.\qquad+\left(\frac{\mu p^+}{6V_5}X_{a_1}-\frac{T_{\text{M5}}}{5!}\epsilon_{a_1\cdots a_6}\left\{X^{a_2},\cdots,X^{a_6}\right\}\right)^2\right).
\label{light-cone-Ham}
\end{align}
Its classical solution with zero light-cone energy is
\begin{align}
P^A=0,\qquad X^i=0,\qquad X^a=\left(\frac{\mu p^+}{6T_{\text{M5}}V_5}\right)^{1/4} x^a,\label{vacuum}
\end{align} 
where $x^a$ is an embedding of the unit 5-sphere that satisfies
\begin{align}
x^ax^a=1,\qquad
\left\{x^{a_2},\cdots,x^{a_5}\right\}=\epsilon^{a_1\cdots a_6}x_{a_1}.
\label{5-bracket-alg}
\end{align}
Therefore, the vacuum configuration of the five-brane theory forms a 5-sphere of radius 
\begin{align}
r_{\text{M5}}=\left(\frac{\mu p^+}{6T_{\text{M5}}V_5}\right)^{1/4}.
\end{align}

\section{Rotating BPS solution}
The Hamiltonian system \eqref{light-cone-Ham} has other classical solutions
which correspond to a rotating squashed 5-sphere.
Since we are interested in such BPS solutions in particular,
we solve the BPS saturation condition in this section.

The supersymmetry algebra of the five-brane theory is
\begin{align}
\left\{Q_I,Q_J\right\}= 2\left(\gamma^\mu\right)_{IJ}P_\mu-\frac{\mu}{3}\left(\gamma^{+123}\gamma^{ij}\right)_{IJ}M_{ij}+\frac{\mu}{6}\left(\gamma^{+123}\gamma^{ab}\right)_{IJ} M_{ab},\label{susy-alg}
\end{align}
where $P_\mu$ and $M_{\mu\nu}$ are the generators of the 11-dimensional Poincar\'e group, 
and $\gamma^A$ are $32 \times 32$ gamma matrices with $I,J$ being the spinor indices.
Therefore, 
the BPS condition can be written as
\begin{align}
0=H+\frac{\eta_1 \mu}{3}M_{12}+\frac{\eta_4 \mu}{6}M_{45}+\frac{\eta_6 \mu}{6}M_{67}+ \frac{\eta_8 \mu}{6}M_{89}, \label{m5bps2}
\end{align}
where $\eta_1,\eta_4,\eta_6,\eta_8=\pm 1$ and $H=P^0$ is the Hamiltonian.
In terms of this condition, 
the static spherical solution in the previous section, 
which is the vacuum configuration,
is a solution to
$H=0$, $M_{ij}=0$, $M_{ab}=0$.

To obtain a non-trivial BPS solution,
we rewrite the BPS condition in terms of a summation of squares as
\begin{align}
H+\frac{\eta_1 \mu}{3}M_{12}+\frac{\eta_4 \mu}{6}M_{45}+\frac{\eta_6 \mu}{6}M_{67}+\frac{\eta_8 \mu}{6}M_{89}=\hana H_1+\hana H_2 +\hana H_3+\hana H_4, \label{m5bpsH}
\end{align}
where $\hana H_1$ is the part that contains momenta\footnote{
In rewriting the BPS condition,
we introduce new parameters $\hat\eta_a=\pm 1$ ($a=4,6,8$), which will be determined shortly.
}
\begin{align*}
\hana H_1 =&\frac{V_5}{2p^+}\int d^5\sigma \bigg[\bigg(P^1-\eta_1\frac{\mu p^+}{3V_5}X^2-\sum_{(a,b,c)}\eta_a N_{1b(b+1)c(c+1)}\bigg)^2 
+\bigg(P^2+\eta_1\frac{\mu p^+}{3V_5}X^1-\sum_{(a,b,c)}\eta_a N_{2b(b+1)c(c+1)}\bigg)^2\no
&+\bigg(P^3-\sum_{(a,b,c)}\eta_a N_{3b(b+1)c(c+1)}-\sum_{(a,b,c)}\hat \eta_a N_{312a(a+1)}\bigg)^2+\bigg(\frac{\mu p^+}{3V_5}\bigg)^2X^3X^3\no
&+ \sum_{(a,b,c)}\bigg\{\bigg(P^{a}-\eta_a\frac{\mu p^+}{6V_5}X^{a+1}-\eta_a N_{ab(b+1)c(c+1)}-\hat \eta_b N_{a12b(b+1)}-\hat \eta_c N_{a12c(c+1)}\bigg)^2 \no
&\qquad+\bigg(P^{a+1}+\eta_a\frac{\mu p^+}{6V_5}X^{a}-\eta_a N_{(a+1)b(b+1)c(c+1)}-\hat \eta_b N_{(a+1)12b(b+1)}-\hat \eta_c N_{(a+1)12c(c+1)}\bigg)^2\bigg\}\bigg],
\end{align*}
$\hana H_2$ is the part that consists only of 5-brackets that do not contain $X^3$
\begin{align*}
\hana H_2=&\frac{V_5}{2p^+}\int d^5\sigma \bigg[\bigg(\eta_1N_{12468}+\sum_{(a,b,c)}\hat \eta_a N_{12a(b+1)(c+1)}\bigg)^2+\bigg(\eta_1 N_{12579}+\sum_{(a,b,c)}\hat \eta_a N_{12(a+1)bc}\bigg)^2 \no
&+\sum_{(a,b,c)}\bigg(\eta_1N_{1a(a+1)bc}+\hat \eta_a N_{1a(a+1)(b+1)(c+1)}+\eta_b N_{2a(a+1)(b+1)c} +\eta_c N_{2a(a+1)b(c+1)}\bigg)^2\no
&+\sum_{(a,b,c)}\bigg(\eta_1 N_{2a(a+1)(b+1)(c+1)}+\hat \eta_a N_{2a(a+1)bc}+\eta_b N_{1a(a+1)b(c+1)}+\eta_c N_{1a(a+1)(b+1)c}\bigg)^2\bigg],
\end{align*}
$\hana H_3$ is the part that consists of 5-brackets that contain $X^3$
\begin{align*}
\hana H_3 =&\frac{V_5}{2p^+}\int d^5\sigma \bigg[\sum_{i=1,2}\bigg(\eta_1N_{3i468}+\sum_{(a,b,c)}\hat \eta_a N_{3ia(b+1)(c+1)}\bigg)^2+\sum_{i=1,2}\bigg(\eta_1 N_{3i579}+\sum_{(a,b,c)}\hat \eta_a N_{3i(a+1)bc}\bigg)^2 \no
&+\sum_{(a,b,c)}\bigg(\hat \eta_a N_{31ca(a+1)}+\hat \eta_b N_{31cb(b+1)}-\eta_a N_{32(c+1)b(b+1)}-\eta_bN_{32(c+1)a(a+1)}\bigg)^2\no
&+\sum_{(a,b,c)}\bigg(\hat \eta_a N_{31(c+1)a(a+1)}+\hat \eta_b N_{31(c+1)b(b+1)}+\eta_a N_{32cb(b+1)}+\eta_b N_{32ca(a+1)}\bigg)^2\no
&+\sum_{(a,b,c)}\bigg(\bigg(\hat \eta_a N_{312ab}-\hat \eta_b N_{312(a+1)(b+1)}\bigg)^2+\bigg(\hat \eta_a N_{312a(b+1)}+\hat \eta_b N_{312(a+1)b}\bigg)^2\bigg)\no
&+\sum_{(a,b,c)}\bigg(\bigg(\eta_a N_{3c(c+1)ab}-\eta_b N_{3c(c+1)(a+1)(b+1)}\bigg)^2+\bigg(\eta_a N_{3c(c+1)a(b+1)}+\eta_b N_{3c(c+1)(a+1)b}\bigg)^2\bigg)\bigg],
\end{align*}
and $\hana H_4$ is the rest, which is not written as a sum of squares
but has terms proportional to the Gauss law constraint
\begin{align}
 0&=d\sigma^\alpha \wedge d\sigma^\beta\, \partial_{\alpha} P^A \partial_{\beta}X_A ,
 \label{Gausslaw}
\end{align}
and also terms that vanish by a particular choice of $\hat\eta_a$,
\begin{align}
\hana H_4&= \frac{V_5}{p^+}\int d^5\sigma \sum_{(a,b,c)}\bigg[P^A \bigg(\eta_a N_{Ab(b+1)c(c+1)}+\hat \eta_a N_{A12a(a+1)}\bigg)
\nonumber \\
&\hspace{100pt}
+\frac{\mu p^+}{3V_5}X^1\bigg(2\eta_1\eta_a+\eta_b\hat \eta_c+\eta_c\hat \eta_b\bigg)N_{2b(b+1)c(c+1)}\bigg].
\end{align}
In the expressions of $\hana H_1,\cdots,\hana H_4$,
we used the following short-hand notations
\begin{align}
 \sum_{(a,b,c)}=\sum_{(a,b,c)=(4,6,8),(6,8,4),(8,4,6)}.
\end{align}
In addition, we defined
\begin{align}
 N_{A_1\cdots A_5}=T_{\text{M5}}\left\{X_{A_1},\cdots,X_{A_5}\right\}
 ,
\end{align}
and 
used a formula obtained from the fundamental identity of the 5-bracket,
\begin{align}
\int d^5\sigma N_{A_1\cdots A_5}N_{B_1\cdots B_5}&=\int d^5\sigma\Big(
N_{B_1A_2A_3A_4 A_5}N_{A_1B_2B_3B_4 B_5}+N_{A_1B_1A_3A_4 A_5}N_{A_2B_2B_3B_4 B_5}\no
&\qquad\qquad +N_{A_1A_2B_1A_4 A_5}N_{A_3B_2B_3B_4 B_5}+N_{A_1A_2A_3B_1 A_5}N_{A_4B_2B_3B_4 B_5}\no
&\qquad\qquad +N_{A_1A_2A_3A_4 B_1}N_{A_5B_2B_3B_4 B_5}
\Big) .
\end{align}
We set the signatures $\hat\eta_a$ ($a=4,6,8$),
which were introduced in writing $\hana H_1,\cdots \hana H_4$,
to be determined by 
\begin{align}
 0=2\eta_1\eta_a+\eta_b\hat \eta_c+\eta_c\hat \eta_b,
 \label{eta}
\end{align}
which results in
\begin{align}
 \hat\eta_a=-\eta_1\eta_b\eta_c
\end{align}
for $(a,b,c)=(4,6,8), (6,8,4), (8,4,6)$,
so that the second term of $\hana H_4$ vanishes.
Then, under the Gauss law constraint,
the BPS condition (\ref{m5bps2}) 
is written only by perfect squares, 
and thus each square in 
$\hana H_1$, $\hana H_2$ and $\hana H_3$
should vanish.

One immediately finds from the condition $\hana H_1=0$ that
\begin{align}
 X^3=0.
\end{align}
Then, $\hana H_3=0$ is automatic.
The rest of the condition $\hana H_1=0$
is rewritten as follows:
\begin{align}
\partial_0 X^1&=\frac{\eta_1\mu}{3}X^2+\frac{V_5}{p^+}\left(\eta_4 N_{16789}+\eta_6N_{14589}+\eta_8 N_{14567}\right),
\label{H1_BPSeq_1} \\
\partial_0 X^2&=-\frac{\eta_1\mu}{3}X^1+\frac{V_5}{p^+}\left(\eta_4 N_{26789}+\eta_6N_{24589}+\eta_8 N_{24567}\right),
\label{H1_BPSeq_2} \\
\partial_0X^a&=\frac{\eta_a\mu}{6}X^{a+1}+\frac{V_5}{p^+}\left(\eta_a N_{ab(b+1)c(c+1)}+\hat \eta_b N_{a12b(b+1)}+\hat \eta_c N_{a12c(c+1)}\right),
\label{H1_BPSeq_a} \\
\partial_0X^{a+1}&=-\frac{\eta_a\mu}{6}X^{a}+\frac{V_5}{p^+}\left(\eta_a N_{(a+1)b(b+1)c(c+1)}+\hat \eta_b N_{(a+1)12b(b+1)}+\hat \eta_c N_{(a+1)12c(c+1)}\right),
\label{H1_BPSeq_a+1}
\end{align}
for $(a,b,c)=(4,6,8), (6,8,4), (8,4,6)$,
where we use 
\begin{align}
 P^A=\frac{p^+}{V_5}\partial_0 X^A
\end{align}
by gauge-fixing $\lambda^\alpha=0$ 
so that the covariant derivative becomes just a partial differentiation.
The remaining BPS condition $\hana H_2=0$ leads to  
\begin{align}
0&=\eta_1 N_{12468}+\hat \eta_4N_{12479}+\hat \eta_6N_{12569}+\hat \eta_8N_{12578},
\label{H2_BPSeq_1}\\
0&=\eta_1N_{12579}+\hat \eta_4N_{12568}+\hat \eta_6N_{12478}+\hat \eta_8N_{12469},
\label{H2_BPSeq_2}\\
0&=\eta_1N_{1a(a+1)bc}+\hat \eta_a N_{1a(a+1)(b+1)(c+1)}+\eta_b N_{2a(a+1)(b+1)c}+\eta_c N_{2a(a+1)b(c+1)},
\label{H2_BPSeq_3}\\
0&=\eta_1 N_{2a(a+1)(b+1)(c+1)}+\hat \eta_a N_{2a(a+1)bc}+\eta_b N_{1a(a+1)b(c+1)}+\eta_c N_{1a(a+1)(b+1)c},
\label{H2_BPSeq_4}
\end{align}
for $(a,b,c)=(4,6,8), (6,8,4), (8,4,6)$.

Let us now apply the following ansatz, 
\begin{align}
 \begin{pmatrix}
  X^1\\
  X^2
 \end{pmatrix}
 =
 \sum_{a=4,6,8}\hat R_a(t)
 \begin{pmatrix}
  x^{a}\\
  x^{a+1}
 \end{pmatrix}
 ,\qquad
 \begin{pmatrix}
  X^a\\
  X^{a+1}
 \end{pmatrix}
 =R_a(t)
 \begin{pmatrix}
  x^a\\
  x^{a+1}
 \end{pmatrix}
 ,
 \label{ansatz1}
\end{align}
where $\hat R_a(t)$ and $R_a(t)$ are $2\times 2$ real matrices depending only on $t=\sigma^0$.

Then, using \eqref{5-bracket-alg},
the BPS equations \eqref{H1_BPSeq_1} and \eqref{H1_BPSeq_2} are reduced to
\begin{align}
 \partial_0\hat R_a &= \hat \alpha \epsilon \hat R_a-\hat \beta_a \hat R_a \epsilon
 \label{eom1}
\end{align}
for $a=4,6,8$.
On the other hand, 
the BPS equations \eqref{H1_BPSeq_a} and \eqref{H1_BPSeq_a+1} are reduced to
\begin{align}
 &\partial_0R_a = \alpha_a \epsilon R_a-\beta_a R_a \epsilon
 \label{eom2}
\end{align}
for $a=4,6,8$, and
\begin{align}
 0=\left(\det R_a\right)\Lambda^T_b\epsilon \hat R_c ,
 \qquad
 0=\left(\det R_a\right)\Lambda^T_c\epsilon \hat R_b
 \label{eom3}
\end{align}
for $(a,b,c)=(4,6,8), (6,8,4), (8,4,6)$,
where we define
$\Lambda_a=\hat R_a \epsilon R^T_a$.
Here,
the coefficients $\hat\alpha$, $\hat\beta_a$, $\alpha_a$ and $\beta_a$ satisfy
\begin{align}
 &\hat \alpha =\frac{\eta_1\mu}{3},
 \qquad
 \hat \beta_a=\frac{\eta_a\mu}{6r^4_{\text{M5}}} \det R_b \det R_c ,
 \nonumber \\
 &\alpha_a = \frac{\eta_a\mu}{6},
 \qquad
 \beta_a
 =\frac{\eta_a\mu}{6r^4_{\text{M5}}}
 \left( \det R_b \det R_c
 -\eta_1 \eta_c \det R_b\det \hat R_c
 -\eta_1 \eta_b \det \hat R_b\det R_c \right)
 ,
\end{align}
for $(a,b,c)=(4,6,8), (6,8,4), (8,4,6)$,
and the matrix $\epsilon$ is the totally antisymmetric tensor $\epsilon=i\sigma_2$,
with $\sigma_i$ being the Pauli matrices, which satisfy $[\sigma_i,\sigma_j]=2i\epsilon_{ijk}\sigma_k$.
It is then straightforward from \eqref{eom1} and \eqref{eom2}
to show that $\det\hat R_a$ and $\det R_a$ are time independent
and thus the coefficients $\hat\alpha$, $\hat\beta_a$, $\alpha_a$ and $\beta_a$ are constant.
Therefore,
the solution to \eqref{eom1} and \eqref{eom2} is
\begin{align}
 \hat R_a(t)=\hana R(- \hat\alpha t)\hat R_a(0)\hana R( \hat\beta_a t),
 \qquad
 R_a(t)= \hana R(- \alpha_a t) R_a(0)\hana R( \beta_a t),
 \label{BPSsolR}
\end{align}
where $\hana R(\theta)=\exp(-\theta\epsilon)=I\cos \theta -\epsilon \sin \theta$
with $I$ being the identity matrix.
Note that the solution implies that
$\tr(\hat R_a^T\hat R_a)$ and $\tr(R_a^TR_a)$ are time independent as well.

The solution \eqref{BPSsolR} needs to satisfy $\hana H_2=0$ and the Gauss law constraint as well.
One can find that
the BPS equations \eqref{H2_BPSeq_1} and \eqref{H2_BPSeq_2} are equivalent to
\begin{align}
 &0
 =\left(
 \eta_b\eta_c\epsilon\Lambda_b^T\epsilon\Lambda_c
 +\Lambda_b^T\epsilon\Lambda_c\epsilon
 \right)
 \begin{pmatrix}
  \eta_c&0\\
  0&-\eta_a
 \end{pmatrix}
 R_a
 ,
 \label{eom4}
\end{align}
and the BPS equations \eqref{H2_BPSeq_3} and \eqref{H2_BPSeq_4} to
\begin{align}
 &0
 =\det R_c
 \left(
 \eta_1\eta_a\epsilon\Lambda_a
 +\Lambda_a\epsilon
 \right)
 \begin{pmatrix}
  \eta_a&0\\
  0&-\eta_b
 \end{pmatrix}
 R_b
 , \qquad
 0
 =\det R_b
 \left(
 \eta_1\eta_a\epsilon\Lambda_a
 +\Lambda_a\epsilon
 \right)
 \begin{pmatrix}
  \eta_a&0\\
  0&-\eta_c
 \end{pmatrix}
 R_c
 \label{eom5}
 ,
\end{align}
for $(a,b,c)=(4,6,8), (6,8,4), (8,4,6)$.
The Gauss law constraint \eqref{Gausslaw} is reduced to
\begin{equation}
 0=
 2\alpha_a \det R_a
 -\beta_a \tr( R^T_aR_a)
 +
 2\hat\alpha \det \hat R_a
 -\hat\beta_a \tr(\hat R_a^T\hat R_a)
 ,
 \label{Gauss1}
\end{equation}
for $a=4,6,8$, and
\begin{equation}
 0=
 2\hat\alpha \hat R_a^T\epsilon\hat R_b
 -\hat\beta_a \epsilon\hat R_a^T\hat R_b
 -\hat\beta_b \hat R_a^T\hat R_b\epsilon
 ,
 \label{Gauss2}
\end{equation}
for $(a,b)=(4,6),(6,8),(8,4)$.
See Appendix~\ref{sec:A} for detailed computations.

Let us now assume $\det R_a$ are all non-zero
since we are interested in dynamics of the five-brane 
extending to the $SO(6)$ directions
(See Appendix \ref{sec:B} for the other cases).
In this case, 
the condition \eqref{eom5} for $\hana H_2=0$ is equivalent to 
\begin{align}
 \eta_1\eta_a\epsilon\Lambda_a=-\Lambda_a\epsilon
 ,
 \label{BPSH2}
\end{align}
for $a=4,6,8$.
This implies
\begin{align}
 \eta_1\eta_a(\det\hat R_a)\, R_a^TR_a=-(\det R_a)\, \hat R_a^T\hat R_a,
 \label{BPSbracket}
\end{align}
using the identity $\epsilon M^T\epsilon M=-\det M\, I$, which holds for any $2\times 2$ matrix $M$.
Incidentally, one finds from the BPS equation \eqref{eom3} that 
at least two out of three $\det\hat R_a$ are zero.
Then, say $\det\hat R_6=\det\hat R_8=0$, 
one obtains from \eqref{BPSbracket} that 
$\hat R_6^T\hat R_6=0$ and $\hat R_8^T\hat R_8=0$,
meaning $\hat R_6=\hat R_8=0$.
Therefore, at least two of $\hat R_a$ can be set to zero
without loss of generality.
Note that the BPS equations \eqref{eom3}, \eqref{eom4} and \eqref{Gauss2} are
now automatic.
In the following, we set 
\begin{equation}
 \hat R_6=\hat R_8=0
\end{equation}
and denote $\hat R_4$ just by $\hat R$.

Then, introducing
a constant $A$ defined by
\begin{align}
 A=\frac{\beta_4 \tr\left(R^T_4R_4\right)}{2\alpha_4 \det R_4}
 ,
 \label{def_of_A}
\end{align}
one can obtain from \eqref{Gauss1} and \eqref{BPSbracket} for $a=4$ that
\begin{align}
 (A+2) \det\hat R
 =\eta_1\eta_4(A-1)\det R_4
 , \label{det_hat-R_det_R4}
\end{align}
using $\alpha_4=\frac{\eta_1\eta_4}{2}\hat\alpha$ and $\hat\beta_4=\beta_4$,
and therefore
\begin{align}
 \beta_6 =\frac{\eta_6\mu}{6r^4_{\text{M5}}} \frac{3}{A+2} \det R_4\det R_8,
 \qquad
 \beta_8 =\frac{\eta_8\mu}{6r^4_{\text{M5}}} \frac{3}{A+2} \det R_4\det R_6.
 \label{beta6_beta8}
\end{align}
While the solution \eqref{BPSsolR} has eight parameters\footnote{
$\hat R(0)$ and $R_a(0)$ have two parameters each  
because two parameters (out of four) in each matrix 
can be eliminated by an $SO(2)\times SO(2)\times SO(2)(\subset SO(6))$ rotation of $(X^a,X^{a+1})$
and redefinition of $x^a$ with an $SO(2)\times SO(2)\times SO(2)(\subset SO(6))$ rotation.
} up to an $SO(6)$ rotation,
the number of constraints the solution satisfies
is one in \eqref{BPSH2} plus three in the Gauss law constraint \eqref{Gauss1}.
Therefore,
the solution \eqref{BPSsolR} is written by four parameters:
$A$, $\det R_4$, $\det R_6$ and $\det R_8$.
These parameters satisfy the following inequalities
so that $\hat R(0)$ and $R_a(0)$ are real matrices:
\begin{align}
 0<A\le 1,
 \quad
 r_{\text{M5}}^4 A \ge |\det R_6 \det R_8|,
 \quad
 r_{\text{M5}}^4\frac{A+2}{3} \ge |\det R_4 \det R_8|,
 \quad
 r_{\text{M5}}^4\frac{A+2}{3} \ge |\det R_4 \det R_6|.
 \label{BPS-param-inequality}
\end{align}
See Appendix \ref{sec:inequalities} for details.

In summary,
the BPS solution we have obtained with non-zero $\det R_a$ is 
the expression \eqref{ansatz1} with 
\begin{align}
 &\hat R(t)=\hana R(- \hat\alpha t)\hat R(0)\hana R( \beta_4 t)
 , \qquad
 R_a(t)= \hana R(- \alpha_a t) R_a(0)\hana R( \beta_a t)
 , 
 \label{BPSsolR_final}
\end{align}
where $\hat R(0)$ and $R_a(0)$ satisfy the conditions \eqref{BPSH2} and \eqref{Gauss1}
\begin{align}
 &\eta_1\eta_4\epsilon\hat R(0) \epsilon R^T_4(0)=-\hat R(0) \epsilon R^T_4(0)\epsilon
 , 
 \label{BPSH2_final}
 \\
 &0=
 2\alpha_a \det R_a
 -\beta_a \tr( R^T_aR_a)
 +\delta_{a4}\left(
 2\hat\alpha \det \hat R
 -\beta_4 \tr(\hat R^T\hat R)
 \right)
 , 
 \label{Gauss1_final}
\end{align}
and also the inequalities \eqref{BPS-param-inequality}.
The angular momentum 
$M_{AB}=\int d^5\sigma(X^AP^B-X^BP^A)$
of the obtained solution is computed,
by using the orthogonality 
$\int d^5\sigma\, x^ax^b=\frac{V_5}{6}\delta^{ab}$,
as follows:
\begin{align}
 M_{12}
 &=-\frac{p^+}{18}\frac{\eta_1\mu}{r^4_{\text{M5}}}\det R_4 \det R_6 \det R_8\frac{1-A}{A+2}
 \left( \frac{2r^8_{\text{M5}}A}{(\det R_6 \det R_8)^2}+1 \right)
 ,
 \nonumber \\
 M_{45}
 &=-\frac{p^+}{18}\frac{\eta_4\mu}{r^4_{\text{M5}}}\det R_4 \det R_6 \det R_8
 \left( \frac{r^8_{\text{M5}}A}{(\det R_6 \det R_8)^2}-1 \right)
 ,
 \nonumber \\
 M_{67}
 &=-\frac{p^+}{18}\frac{\eta_6\mu}{r^4_{\text{M5}}}\det R_4 \det R_6 \det R_8 \frac{3}{A+2}
 \left( \left(\frac{A+2}{3}\right)^2\frac{r^8_{\text{M5}}}{(\det R_4 \det R_8)^2}-1 \right)
 ,
 \nonumber \\
 M_{89}
 &=-\frac{p^+}{18}\frac{\eta_8\mu}{r^4_{\text{M5}}}\det R_4 \det R_6 \det R_8 \frac{3}{A+2}
 \left( \left(\frac{A+2}{3}\right)^2\frac{r^8_{\text{M5}}}{(\det R_4 \det R_6)^2}-1 \right)
 .
 \label{BPS-angular-momentum}
\end{align}
One can find 
from the inequalities \eqref{BPS-param-inequality} that
$\eta_1\mu M_{12}$, $\eta_4\mu M_{45}$,
$\eta_6\mu M_{67}$ and $\eta_8\mu M_{89}$, which appear in \eqref{m5bpsH}, 
are all semi-negative definite.

Note we have obtained a general solution in principle.
From \eqref{BPS-angular-momentum}, 
one can translate four components of the angular momentum
to $A$, $\det R_4$, $\det R_6$ and $\det R_8$.
Then, by solving \eqref{BPSH2_final} and \eqref{Gauss1_final},
one obtains the matrices $\hat R(0)$, $R_4(0)$, $R_6(0)$ and $R_8(0)$
and plugs them into \eqref{BPSsolR_final} to arrive at a BPS solution.
We demonstrate the procedure in the following.

\subsection*{A two-parameter solution with $M_{67}=M_{89}=0$}
After a straightforward calculation,
we find there is no solution to \eqref{BPS-angular-momentum} 
if we set $M_{12}\ne 0$ and
three other components of the angular momentum to zero.
We thus consider the case
of two parameters $M_{12}$ and $M_{45}$, 
with $M_{67}=M_{89}=0$, as a simplest non-trivial example.
As we will see later,
a solution with $M_{12}=0$ is included in the case as the one with $A=1$.

To obtain the solution explicitly,
we expand $\hat R(0)$ and $R_a(0)$ 
in terms of the identity matrix $I$, $\epsilon=i\sigma_2$ and the real Pauli matrices $\sigma_1$, $\sigma_3$.
By the $SO(6)$ symmetry for $X^a$ and redefinition of $x^a$ with an $SO(6)$ rotation,
one can simplify them as
\begin{align}
 \hat R(0)
 =\hat c_{0}I
 +\hat c_{3}\sigma_3,
 \qquad
 R_a(0)
 =c_{a0}I
 + c_{a3}\sigma_3,
 \label{initial2}
\end{align}
for $a=4,6,8$,
where $\hat c_{0}$ and $c_{a0}$ are set to be positive.

In this case, where $M_{67}=M_{89}=0$, we have
$\beta_6=\pm\alpha_6$ and 
$\beta_8=\pm\alpha_8$.
Then, by the Gauss law constraint \eqref{Gauss1_final},
one can show that $c_{a3}=0$ if $\beta_a=+\alpha_a$
and that $c_{a0}=0$ if $\beta_a=-\alpha_a$, for $a=6,8$.
In either case, $R_a$ ($a=6,8$) becomes time independent,
proportional to the identity matrix $I$ or $\sigma_3$.
By choosing $\beta_6=\alpha_6$ and $\beta_8=\alpha_8$ for simplicity, 
we obtain
\begin{align}
 R_6(t)=r'_{\text{M5}}I,\qquad R_8(t)=r'_{\text{M5}}I,
 \label{non-rotating_67_89}
\end{align}
where $r'_{\text{M5}}$ is a positive constant parameter.

The BPS condition \eqref{BPSH2_final} is now rewritten as
\begin{align}
 0=(1+\eta_1\eta_4)(\hat c_{0}c_{40}-\hat c_{3}c_{43})I
 -(1-\eta_1\eta_4)(\hat c_{0}c_{43}-\hat c_{3}c_{40})\sigma_3.
\end{align}
As already seen, the Gauss law constraint \eqref{Gauss1_final} for $a=4$
is equivalent to \eqref{det_hat-R_det_R4}.
Using these conditions 
in addition to 
the other Gauss law constraint $\beta_6=\alpha_6$ with \eqref{beta6_beta8} 
and the definition \eqref{def_of_A} of $A$,
we rewrite the parameters $\hat c_{0}$, $\hat c_{3}$, $c_{40}$ and $c_{43}$
in terms of $r'_\text{M5}$ and $A$
and arrive at the following solution:
\begin{align}
 R_4(t)
 &=\frac{r^4_{\text{M5}}}{r'^3_{\text{M5}}}
 \left(\frac{A+2}{6}\left(A+\frac{r'^4_{\text{M5}}}{r^4_{\text{M5}}}\right)\right)^{1/2}
 \hana R\left( -\tfrac{\eta_4\mu}{6}(1-\tfrac{r'^4_{\text{M5}}}{r^4_{\text{M5}}}) t \right) \, I\no
 &\qquad
 \pm \frac{r^4_{\text{M5}}}{r'^3_{\text{M5}}}
 \left(\frac{A+2}{6}\left(A-\frac{r'^4_{\text{M5}}}{r^4_{\text{M5}}}\right)\right)^{1/2}
 \hana R\left( -\tfrac{\eta_4\mu}{6}(1+\tfrac{r'^4_{\text{M5}}}{r^4_{\text{M5}}}) t \right) \, \sigma_3
 ,
 \no
 \hat R(t)
 &=\frac{r^4_{\text{M5}}}{r'^3_{\text{M5}}}
 \left( \frac{1-A}{6}\left(A-\eta_1\eta_4\frac{r'^4_{\text{M5}}}{r^4_{\text{M5}}}\right) \right)^{1/2}
 \hana R\left( -\tfrac{\eta_4\mu}{6}(2\eta_1\eta_4-\tfrac{r'^4_{\text{M5}}}{r^4_{\text{M5}}}) t \right) \, I\no
 &\qquad
 \pm\frac{r^4_{\text{M5}}}{r'^3_{\text{M5}}}
 \left( \frac{1-A}{6}\left(A+\eta_1\eta_4\frac{r'^4_{\text{M5}}}{r^4_{\text{M5}}}\right) \right)^{1/2}
 \hana R\left( -\tfrac{\eta_4\mu}{6}(2\eta_1\eta_4+\tfrac{r'^4_{\text{M5}}}{r^4_{\text{M5}}}) t \right) \, \sigma_3
 .
 \label{result_R}
\end{align}
Here the signs $\pm$ are correlated,
and the same applies to the following expressions.
The solution is then rewritten as
\begin{align}
 &
 \begin{pmatrix}
  X^4\\
  X^5
 \end{pmatrix}
 =\frac{r^4_\text{M5}}{r'^3_\text{M5}}\sqrt{\frac{A+2}{6}}
 \mathscr{R}(-\tfrac{\eta_4\mu}{6}t)
 \left( \sqrt{A+\frac{r'^4_\text{M5}}{r^4_\text{M5}}}\, I
 \pm\sqrt{A-\frac{r'^4_\text{M5}}{r^4_\text{M5}}}\,\sigma_3 \right)
 \mathscr{R}(\tfrac{\eta_4\mu}{6}\tfrac{r'^4_\text{M5}}{r^4_\text{M5}}t)
 \begin{pmatrix}
  x^4\\
  x^5
 \end{pmatrix}
 , \label{result45} \\
 &
 \begin{pmatrix}
  X^1\\
  X^2
 \end{pmatrix}
 =\frac{r^4_\text{M5}}{r'^3_\text{M5}}\sqrt{\frac{1-A}{6}}
 \mathscr{R}(-\tfrac{\eta_1\mu}{3}t)
 \left( \sqrt{A-\eta_1\eta_4\frac{r'^4_\text{M5}}{r^4_\text{M5}}}\, I
 \pm\sqrt{A+\eta_1\eta_4\frac{r'^4_\text{M5}}{r^4_\text{M5}}}\,\sigma_3 \right)
 \mathscr{R}(\tfrac{\eta_4\mu}{6}\tfrac{r'^4_\text{M5}}{r^4_\text{M5}}t)
 \begin{pmatrix}
  x^4\\
  x^5
 \end{pmatrix}
 .
 \label{result12}
\end{align}

The configuration of this BPS solution 
describes an ellipsoidal five-brane rotating
without changing its shape. 
Using the following rotated coordinates
\begin{align}
 \begin{pmatrix}
  \tilde X^4(t)\\
  \tilde X^5(t)
 \end{pmatrix}
 =\mathscr{R}(\tfrac{\eta_4\mu}{6}t)
 \begin{pmatrix}
  X^4(t)\\
  X^5(t)
 \end{pmatrix}
 ,\qquad
 \begin{pmatrix}
  \tilde X^1(t)\\
  \tilde X^2(t)
 \end{pmatrix}
 =\mathscr{R}(\tfrac{\eta_1\mu}{3}t)
 \begin{pmatrix}
  X^1(t)\\
  X^2(t)
 \end{pmatrix}
 ,
\end{align}
one can describe the configuration \eqref{result45} and \eqref{result12} as
\begin{align}
 &\frac{(\tilde X^4)^2}{1\pm\sqrt{1-\frac{r'^8_\text{M5}}{A^2r^8_\text{M5}}}}
 +\frac{(\tilde X^5)^2}{1\mp\sqrt{1-\frac{r'^8_\text{M5}}{A^2r^8_\text{M5}}}}
 =\frac{r^8_\text{M5}}{r'^6_\text{M5}}\frac{A(A+2)}{3}
 (x_4^2+x_5^2)
 ,
 \nonumber \\
 &\frac{(\tilde X^1)^2}{1\pm\sqrt{1-\frac{r'^8_\text{M5}}{A^2r^8_\text{M5}}}}
 +\frac{(\tilde X^2)^2}{1\mp\sqrt{1-\frac{r'^8_\text{M5}}{A^2r^8_\text{M5}}}}
 =\frac{r^8_\text{M5}}{r'^6_\text{M5}}\frac{A(1-A)}{3}
 (x_4^2+x_5^2)
 .
\end{align}
Hence, the shape of the configuration is ellipsoidal and time-independent.
Note that the angular velocity of the five-dimensional ellipsoid is 
$-\frac{\eta_4\mu}{6}$ on the $(X^4,X^5)$-plane and 
$-\frac{\eta_1\mu}{3}$ on the $(X^1,X^2)$-plane,
which are independent of $r'_\text{M5}$ and $A$,
while the angular momentum $M_{AB}$ depends on the parameters.

\begin{figure}[htbp]
\centering
  \includegraphics[width=16.0cm]{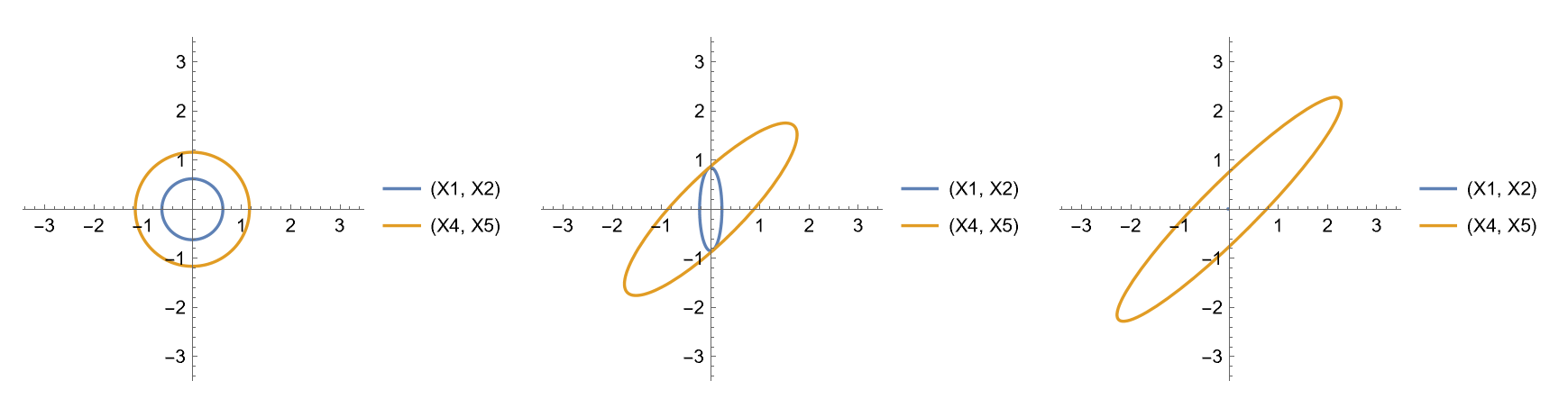}
 \caption{\small $A$-dependence of the curves $(X^1(\sigma),X^2(\sigma))$ and $(X^4(\sigma),X^5(\sigma))$ at $t=\pi/4$ with $r'^4_{\text{M5}}/r^4_{\text{M5}}$ fixed to $1/3$. The left, center and right graphs depict the cases of $A=1/3$, $A=2/3$ and $A=1$, respectively. We set $\mu=-6\eta_4$ and $\eta_1=\eta_4$. One can see that the $(X^1,X^2)$ direction shrinks to zero in the right-most graph ($A=1$).}
  \label{No1}
\end{figure}

The solution is written only by two parameters, $r'_{\text{M5}}$ and $A$,
which are restricted to the range 
\begin{align}
0< \frac{r'^4_{\text{M5}}}{r^4_{\text{M5}}}\le A \le 1.
\end{align}
The parameters $r'_{\text{M5}}$ and $A$ determine
not only the angular momentum of the five-brane,
but also its size and shape.
In particular, when $A=1$, 
all of the $SO(3)$ directions $X^i$ vanish and 
the remaining six directions $X^a$ describe the five-brane. 
See Fig.~\ref{No1} for depiction of $A$-dependence of the curves $(X^1,X^2)$ and $(X^4,X^5)$.

\begin{figure}[htbp]
\centering
  \includegraphics[width=16.0cm]{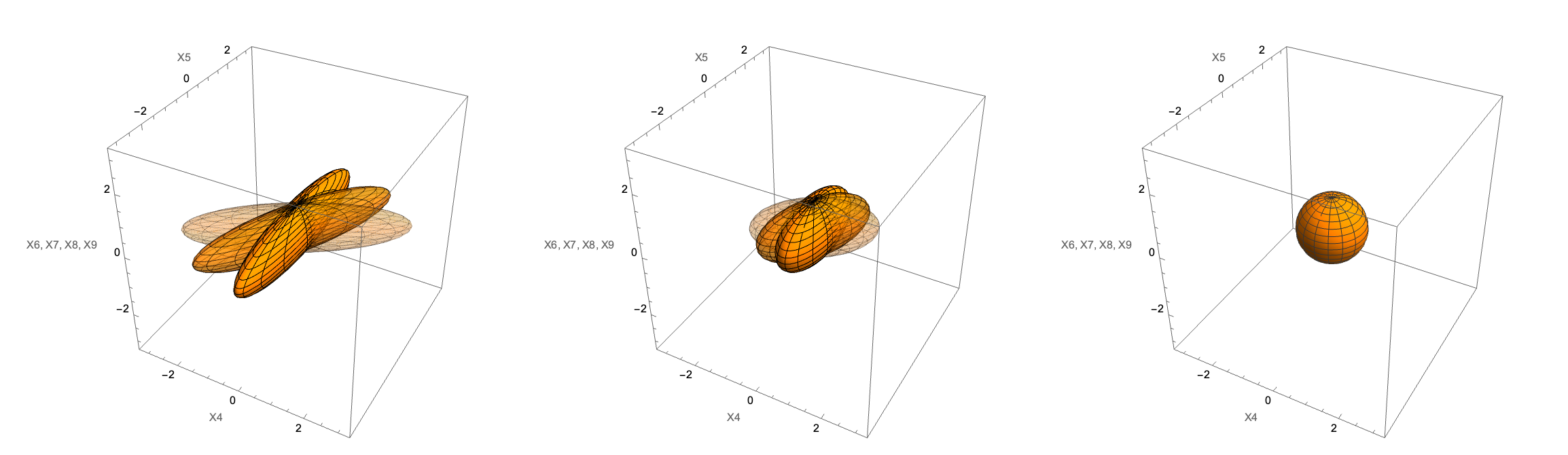}
  \caption{\small $r'_{\text{M5}}$-dependence of the shape of the rotating five-brane with $A=1$. The left, center and right figures show the cases of $r'^4_{\text{M5}}/r^4_{\text{M5}}=1/3$, $r'^4_{\text{M5}}/r^4_{\text{M5}}=2/3$ and $r'^4_{\text{M5}}/r^4_{\text{M5}}=1$, respectively. We set $\mu=-6\eta_4$. One can see that the brane becomes $S^5$ in the right-most figure. The time dependence is indicated by different transparencies: the more transparent, the earlier.}
  \label{No2}
\end{figure}

Note also that
the static spherical solution, seen in section 2, 
is contained in the BPS solution 
as the solution with $A=r'_{\text{M5}}/r_{\text{M5}}=1$:
\begin{equation}
R_a(t)=r_{\text{M5}} I,
\qquad
\hat R(t)=0,
\end{equation}
It is shown in Fig.~\ref{No2} that
the configuration with $A=1$ 
tends to the spherical solution
as $r'_{\text{M5}}/r_{\text{M5}}$ approaches $1$,
rotating as time $t$ varies
without changing its shape
before $r'_{\text{M5}}/r_{\text{M5}}$ reaches $1$.


\section{Summary}

We derived a family of BPS solutions
in the bosonic sector of the five-brane theory.
In particular, we focused on BPS solutions that correspond to five-brane dynamics
with a finite angular momentum.
To demonstrate the solution,
we considered a simple case with $M_{67}=M_{89}=0$
and presented its explicit form, 
which describes a rotating ellipsoidal five-brane.

Our original motivation was to derive BPS solutions
that correspond to the quarter-BPS sector of the matrix model
studied in \cite{Asano:2017xiy,Asano:2017nxw}.
Since the matrix model is double Wick-rotated in the study,
the corresponding BPS solutions should be 
those in the double Wick-rotated five-brane theory.
Although the obtained BPS solutions are 
the ones without the double Wick rotation,
we expect one can use the methodology developed in this paper
to find the BPS solutions corresponding to the quarter-BPS sector,
which should be described by the fluctuation in the matrix model.

\section*{Acknowledgment}
We thank Hiroyuki Adachi for his contribution at the early stage of this work.
This work was supported by JSPS KAKENHI Grant Numbers JP23K03405 and JP24K07036.

\appendix 
\section{Derivation of BPS conditions}\label{sec:A}
In this appendix,
we rewrite the BPS conditions with the ansatz \eqref{ansatz1}.

\subsection{BPS equations from $\hana H_2=0$} \label{A.2}
Let us compute the BPS equations \eqref{H2_BPSeq_1}--\eqref{H2_BPSeq_4},
which arise from $\hana H_2=0$.


To compute the BPS equations, 
we explicitly write
\begin{align}
 \hat R_a=
 \begin{pmatrix}
  \hat R^{1\, a}&\hat R^{1\, (a+1)}\\
  \hat R^{2\, a}&\hat R^{2\, (a+1)} 
 \end{pmatrix}
 ,\qquad
 R_a=
 \begin{pmatrix}
  R^{a\, a}&R^{a\, (a+1)}\\
  R^{(a+1)\, a}&R^{(a+1)\, (a+1)}
 \end{pmatrix}
 ,
\end{align}
and $\Lambda_a=\hat R_a\epsilon R_a^T$ as
\begin{align}
 \Lambda_a 
 =
 \begin{pmatrix}
  \Lambda^{1\, a} & \Lambda^{1\, (a+1)}\\
  \Lambda^{2\, a} & \Lambda^{2\, (a+1)}
 \end{pmatrix}
 =
 \begin{pmatrix}
  \hat R^{1\, a}R^{a\, (a+1)}-\hat R^{1\, (a+1)}R^{a\, a} & \hat R^{1\, a}R^{(a+1)\, (a+1)}-\hat R^{1\, (a+1)}R^{(a+1)\, a}\\
  \hat R^{2\, a}R^{a\, (a+1)}-\hat R^{2\, (a+1)}R^{a\, a} & \hat R^{2\, a}R^{(a+1)\, (a+1)}-\hat R^{2\, (a+1)}R^{(a+1)\, a}
 \end{pmatrix}
 .
\end{align}
Using this notation, 
one can rewrite 5-brackets 
as follows.
For $(a,b,c)=(4,6,8),(6,8,4),(8,4,6)$,
\begin{align}
 \left\{X^1,X^{a},X^{a+1},X^{B},X^{C}\right\}
 &= (\det R_a)
 \left(
 \Lambda^{1B} \left\{x^{a},x^{a+1},x^{b},x^{b+1},X^{C}\right\}
 -\Lambda^{1C} \left\{x^{a},x^{a+1},x^{c},x^{c+1},X^{B}\right\}
 \right)
 , \nonumber \\
 \left\{X^2,X^{a},X^{a+1},X^{B},X^{C}\right\}
 &= (\det R_a)
 \left(
 \Lambda^{2B} \left\{x^{a},x^{a+1},x^{b},x^{b+1},X^{C}\right\}
 -\Lambda^{2C} \left\{x^{a},x^{a+1},x^{c},x^{c+1},X^{B}\right\}
 \right)
 \label{N1aabc}
 ,
\end{align}
where the index $B$ indicates either $b$ or $b+1$ and $C$ indicates $c$ or $c+1$.
We also have
\begin{align}
 \left\{X^1,X^2,X^A,X^B,X^C\right\}
 &=-\Delta^{AB}\left\{x^a,x^{a+1},x^b,x^{b+1},X^C\right\}
 -\Delta^{BC}\left\{x^b,x^{b+1},x^c,x^{c+1},X^A\right\}
 \nonumber \\
 &\qquad
 -\Delta^{CA}\left\{x^a,x^{a+1},x^c,x^{c+1},X^B\right\}
 \label{N12abc}
 ,
\end{align}
where the index $A$ is either 4 or 5, $B$ is 6 or 7, and $C$ is 8 or 9,
and $\Delta^{bc}$ is defined by
\begin{align}
 \Lambda^T_b\epsilon \Lambda_c
 =&
 \begin{pmatrix}
  \Lambda^{1b}\Lambda^{2c}-\Lambda^{2b}\Lambda^{1c}&\Lambda^{1b}\Lambda^{2(c+1)}-\Lambda^{2b}\Lambda^{1(c+1)}\\
  \Lambda^{1(b+1)}\Lambda^{2c}-\Lambda^{2(b+1)}\Lambda^{1c}&\Lambda^{1(b+1)}\Lambda^{2(c+1)}-\Lambda^{2(b+1)}\Lambda^{1(c+1)}
 \end{pmatrix}
 \equiv 
 \begin{pmatrix}
  \Delta^{bc} & \Delta^{b(c+1)}\\
  \Delta^{b(c+1)}& \Delta^{(b+1)(c+1)}
 \end{pmatrix}  
.
\end{align}

The right-hand side of
the BPS equations \eqref{H2_BPSeq_3} and \eqref{H2_BPSeq_4} is
reduced to
\begin{align}
&\left(\begin{array}{c}
\eta_1N_{1a(a+1)bc}+\hat \eta_a N_{1a(a+1)(b+1)(c+1)}+\eta_c N_{2a(a+1)b(c+1)}+\eta_b N_{2a(a+1)(b+1)c}\\
\eta_1N_{2a(a+1)(b+1)(c+1)}+\hat \eta_a N_{2a(a+1)bc}+\eta_c N_{1a(a+1)(b+1)c}+\eta_b N_{1a(a+1)b(c+1)}
\end{array}\right)\no
&=T_{\text{M5}}\det R_a\left(\begin{array}{cc}
\eta_b &0\\
0&-\eta_c
\end{array}\right)
\left(\eta_1\Lambda_c-\eta_c\epsilon \Lambda_c\epsilon\right)\left(\begin{array}{cc}
\eta_b &0\\
0&-\eta_c
\end{array}\right)R_b\epsilon\left(\begin{array}{c}
x^{b}\\
x^{b+1}
\end{array}
\right)\no
&\qquad
-T_{\text{M5}}\det R_a\left(\begin{array}{cc}
\eta_b &0 \\
0&-\eta_c
\end{array}\right)
\left(\eta_1\Lambda_b-\eta_b\epsilon \Lambda_b \epsilon
\right)\left(\begin{array}{cc}
\eta_b &0 \\
0&-\eta_c
\end{array}\right)R_c\epsilon\left(\begin{array}{c}
x^c\\
x^{c+1}
\end{array}
\right) 
,
\end{align}
using \eqref{N1aabc} and $\hat \eta_a = -\eta_1\eta_b\eta_c$.
Since it holds for $(a,b,c)=(4,6,8),(6,8,4),(8,4,6)$,
we obtain a simpler form of the BPS equations as
\eqref{eom5}:
\begin{align*}
 &0
 =\det R_c
 \left(
 \eta_1\eta_a\epsilon\Lambda_a
 +\Lambda_a\epsilon
 \right)
 \begin{pmatrix}
  \eta_a&0\\
  0&-\eta_b
 \end{pmatrix}
 R_b
 , \qquad
 0
 =\det R_b
 \left(
 \eta_1\eta_a\epsilon\Lambda_a
 +\Lambda_a\epsilon
 \right)
 \begin{pmatrix}
  \eta_a&0\\
  0&-\eta_c
 \end{pmatrix}
 R_c
 .
\end{align*}

For the BPS equations \eqref{H2_BPSeq_1} and \eqref{H2_BPSeq_2},
the right-hand side is reduced to
\begin{align}
&\left(\begin{array}{c}
\eta_1N_{12468}+\hat \eta_4N_{12479}+\hat \eta_6N_{12569}+\hat \eta_8N_{12578}\\
\eta_1N_{12579}+\hat \eta_4N_{12568}+\hat \eta_6N_{12478}+\hat \eta_8N_{12469}
\end{array}\right)\no
&=T_{\text{M5}}\sum_{(a,b,c)}\left(\begin{array}{cc}
\hat \eta_a  &0\\
0&-\hat \eta_c
\end{array}\right)\left(\eta_1\Lambda^T_b\epsilon \Lambda_c-\hat \eta_a\epsilon \Lambda^T_b\epsilon \Lambda_c\epsilon\right)\left(\begin{array}{cc}
\hat \eta_a  &0\\
0&-\hat \eta_c
\end{array}\right)R_a\epsilon
\left(\begin{array}{c}
x^a\\
x^{a+1}
\end{array}
\right)
,
\end{align}
using \eqref{N12abc}.
We thus obtain a simpler form of the BPS equations as
\eqref{eom4}:
\begin{equation*}
 0
 =\left(
 \eta_b\eta_c\epsilon\Lambda_b^T\epsilon\Lambda_c
 +\Lambda_b^T\epsilon\Lambda_c\epsilon
 \right)
 \begin{pmatrix}
  \eta_c&0\\
  0&-\eta_a
 \end{pmatrix}
 R_a
 .
\end{equation*}

\subsection{Gauss law constraint}\label{A.3}
The Gauss law constraint \eqref{Gausslaw} is rewritten as 
\begin{align}
 d\sigma^\alpha \wedge d\sigma^\beta\,
 \partial_{\alpha} X^A \partial_{\beta}\partial_0X_A
 &=
 d\sigma^\alpha \wedge d\sigma^\beta\,
 \sum_{a=4,6,8} \sum_{b=4,6,8}
 \begin{pmatrix}
  \partial_{\alpha} x^{a} & \partial_{\alpha} x^{a+1}
 \end{pmatrix}
 \hat R_a^T \partial_0\hat R_b
 \begin{pmatrix}
  \partial_{\beta} x^{b}\\
  \partial_{\beta} x^{b+1}
 \end{pmatrix}
 \nonumber \\
 &\qquad
 + d\sigma^\alpha \wedge d\sigma^\beta\,
 \sum_{a=4,6,8}
 \begin{pmatrix}
  \partial_{\alpha} x^a & \partial_{\alpha} x^{a+1}
 \end{pmatrix}
 R^T_a \partial_0R_a
 \begin{pmatrix}
  \partial_{\beta} x^a\\
  \partial_{\beta} x^{a+1}
 \end{pmatrix}
 \nonumber \\
 &=
 -\sum_{a=4,6,8}
 \tr\left[
 \epsilon\left(
 R^T_a \partial_0R_a
 +\hat R^T_a \partial_0\hat R_a
 \right) \right]
 dx^a
 \wedge
 dx^{a+1}
 \nonumber \\
 &\qquad
 +
 \sum_{(a,b,c)}
 \begin{pmatrix}
  dx^{a} & dx^{a+1}
 \end{pmatrix}
 \left( \hat R_a^T \partial_0\hat R_b
 -\left( \hat R_b^T \partial_0\hat R_a \right)^T \right)
 \begin{pmatrix}
  dx^{b}\\
  dx^{b+1}
 \end{pmatrix}
 .
\end{align}
For this expression to be zero at any point on the world-volume,
$\hat R_a$ and $R_a$ need to satisfy not only the condition that
each coefficient of $d\sigma^\alpha\wedge d\sigma^\beta$ is zero
but also that each of $dx^a\wedge dx^b$ is zero.
Hence, a set of conditions equivalent to the Gauss law constraint is
\begin{align}
 0 = \tr\left[\epsilon \left(\hat R_a^T \partial_0\hat R_a+R^T_a \partial_0R_a\right)\right]
 , \qquad
 0 = \hat R_a^T \partial_0\hat R_b
 -\left( \hat R_b^T \partial_0\hat R_a \right)^T
 .
\end{align}
By plugging the equation of motion (\ref{eom1}) and (\ref{eom2}) into the conditions, one can obtain \eqref{Gauss1} and \eqref{Gauss2}:
\begin{align*}
 0&=
 2\alpha_a \det R_a
 -\beta_a \tr( R^T_aR_a)
 +
 2\hat \alpha \det \hat R_a
 -\hat \beta_a \tr(\hat R_a^T\hat R_a)
 ,
 \\
 0&=
 2\hat\alpha \hat R_a^T\epsilon\hat R_b
 -\hat\beta_a \epsilon\hat R_a^T\hat R_b
 -\hat\beta_b \hat R_a^T\hat R_b\epsilon
 .
\end{align*}

\section{BPS solutions when at least one of $\det R_a$ is zero}\label{sec:B}
\subsection{$\det R_4=0$}
We discuss the case where only two of $\det R_a$ are non-zero.
Because of the $SO(6)$ symmetry,
we can set
\begin{align}
 \det R_4=0,\qquad
 \det R_6\neq 0,\qquad
 \det R_8\neq 0,
\end{align}
without loss of generality.
In this case, \eqref{BPSH2} for $a=4$ is derived from \eqref{eom5},
and thus
\begin{align}
 \eta_1\eta_4(\det\hat R_4)\, R_4^TR_4
 =0
 \label{BPSbracket_B1}
 .
\end{align}

If $\det\hat R_4=0$, 
then $\hat\beta_6$, $\hat\beta_8$, $\beta_6$ and $\beta_8$ become zero
and hence the Gauss law constraint (\ref{Gauss1}) provides
\begin{align}
 0=2\alpha_6\det R_6
 +2\hat\alpha\det\hat R_6
 ,
 \qquad
 0=2\alpha_8\det R_8
 +2\hat\alpha\det\hat R_8
 ,
\end{align}
meaning $\det\hat R_6$ and $\det\hat R_8$ are non-zero.
However,
these constraints violate the other Gauss law constraint \eqref{Gauss2}.

On the other hand, if $\det\hat R_4\ne 0$, 
then $R_4$ is zero because of the relation \eqref{BPSbracket_B1}.
In addition, $\hat R_6$ and $\hat R_8$ are also zero
because 
$0=R_6\epsilon\hat R_6^T\epsilon\hat R_4=R_8\epsilon\hat R_8^T\epsilon\hat R_4$
from \eqref{eom3}.
Then, the Gauss law constraint \eqref{Gauss1} becomes
\begin{align}
0&= 2\hat \alpha \det \hat R_4 -\hat \beta_4 \tr\left(\hat R_4^T \hat R_4\right), \no
0&= 2\alpha_6 \det R_6- \beta_6 \tr\left( R^T_6 R_6\right), \no
0&= 2\alpha_8 \det R_8-\beta_8 \tr\left(R^T_8 R_8\right).
\end{align}
Although
$\tr(\hat R_4^T\hat R_4)$, $\tr(R^T_6 R_6)$ and $\tr(R^T_8 R_8)$ 
all need to be positive,
as seen in \eqref{matrix-inequality},
this is impossible because
\begin{align}
 &\tr\left(\hat R_4^T \hat R_4\right)
 =\eta_1\eta_4\frac{4r^4_\text{M5}\det \hat R_4}{\det R_6 \det R_8}
 ,
 \nonumber \\
 &\tr\left( R^T_6 R_6\right)
 =-\eta_1\eta_4\frac{2r^4_\text{M5}\det R_6}{\det\hat R_4 \det R_8}
 ,
 \nonumber \\
 &\tr\left(R^T_8 R_8\right)
 =-\eta_1\eta_4\frac{2r^4_\text{M5}\det R_8}{\det\hat R_4 \det R_6}
 ,
\end{align}
where $\tr(\hat R_4^T \hat R_4)$ has the sign opposite
to $\tr( R^T_6 R_6)$ and $\tr( R^T_8 R_8)$.

Therefore, there is no solution in this case.

\subsection{$\det R_6=\det R_8=0$}
We discuss the case where only two of $\det R_a$ are zero;
we set 
\begin{align}
 \det R_4\neq 0,\qquad\det R_6= 0,\qquad\det R_8=0
 \label{caseB3}
\end{align}
without loss of generality.
In this case, $\hat\beta_4$, $\hat\beta_6$, $\hat\beta_8$ and $\beta_4$ are zero.
Then, the Gauss law constraint \eqref{Gauss1} for $a=4$
becomes
\begin{align}
 2\eta_1\det \hat R_4=-\eta_4 \det R_4
 , \label{gaussB3}
\end{align} 
which implies $\det\hat R_4\ne 0$,
and \eqref{Gauss2} for $(a,b)=(4,6),(8,4)$ becomes
\begin{align}
 0=2\hat\alpha\hat R_4^T\epsilon\hat R_6,
 \qquad
 0=2\hat\alpha\hat R_4^T\epsilon\hat R_8,
\end{align}
which leads to $\hat R_6=\hat R_8=0$.
Hence, $\hat\beta_6$ and $\hat\beta_8$ also turn out to be zero,
and one can see all the BPS conditions are satisfied
if \eqref{gaussB3} holds.
Therefore, the solution is 
\begin{align}
 \hat R_4(t)
 \equiv \hat R(t)
 = \hana R\left(-\frac{\mu\eta_1}{3}t\right)\hat R(0),
 \qquad
 R_a(t)= \hana R\left(-\frac{\mu\eta_a}{6}t\right)R_a(0),
\end{align}
with the constraints \eqref{caseB3} and \eqref{gaussB3}.
The angular momentum satisfies
\begin{align}
\frac{\mu\eta_1}{3}M_{12}&=-\frac{p^+}{6}\frac{\mu^2}{9}\tr\left(\hat R^T\hat R\right)<0,\no
\frac{\mu\eta_4}{6}M_{45}&=-\frac{p^+}{6}\frac{\mu^2}{36}\tr\left(R^T_4R_4\right)<0,\no
\frac{\mu\eta_6}{6}M_{67}&=-\frac{p^+}{6}\frac{\mu^2}{36}\tr\left(R^T_6R_6\right)\le 0,\no
\frac{\mu\eta_8}{6}M_{89}&=-\frac{p^+}{6}\frac{\mu^2}{36}\tr\left(R^T_8R_8\right)\le 0.
\end{align}

\subsection{$\det R_4=\det R_6=\det R_8=0$}
We discuss the case where $\det R_a$ are all zero. 
In this case, 
the BPS conditions \eqref{eom3} and \eqref{eom5} are automatically satisfied. 
Since all $\hat\beta_a$ and $\beta_a$ are zero,
the Gauss law constraints (\ref{Gauss1}) and \eqref{Gauss2} are reduced to
$\det\hat R_a=0$ for $a=4,6,8$,
and 
$\hat R_a^T\epsilon\hat R_b=0$ for $(a,b)=(4,6),(6,8),(8,4)$, respectively.
One then sees the BPS condition \eqref{eom4} is satisfied.
Therefore, we arrive at the following solution:
\begin{align}
 \hat R_a (t)= \hana R\left(-\frac{\eta_1\mu}{3}t\right)\hat R_a(0),
 \qquad
 R_a (t)= \hana R\left(-\frac{\eta_a\mu}{6}t\right)R_a(0),
\end{align}
that satisfies 
\begin{align}
 \det\hat R_a=\det R_a=0,
 \qquad
 \hat R_a^T(0)\epsilon\hat R_b(0)=0,
\end{align}
for $(a,b)=(4,6),(6,8),(8,4)$.
The angular momentum 
satisfies
\begin{align}
\frac{\mu\eta_1}{3}M_{12}&=-\frac{p^+}{6}\frac{\mu^2}{9}\sum_{a=4,6,8}\tr\left(\hat R_a^T\hat R_a\right)\le 0,\no
\frac{\mu\eta_4}{6}M_{45}&=- \frac{p^+}{6}\frac{\mu^2}{36}\tr\left(R^T_4R_4\right)\le 0,\no
\frac{\mu\eta_6}{6}M_{67}&=- \frac{p^+}{6}\frac{\mu^2}{36}\tr\left(R^T_6R_6\right)\le 0,\no
\frac{\mu\eta_8}{6}M_{89}&=- \frac{p^+}{6}\frac{\mu^2}{36}\tr\left(R^T_8R_8\right)\le 0.
\end{align}

\section{Inequalities for the parameters}\label{sec:inequalities}
In this appendix,
we derive inequality conditions for $\hat R$ and $R_a$.

Note first that, 
using the explicit form \eqref{initial2} of $\hat R(0)$ and $R_a(0)$, 
one can write down
$\det \hat R$, $\det R_a$, $\tr(\hat R^T \hat R)$ and $\tr( R_a^T R_a)$
as
\begin{alignat}{2}
 &\det \hat R
 = {\hat c_{0}}^2
 -{\hat c_{3}}^2,
 &\qquad &
 \frac{1}{2}\tr\left(\hat R^T\hat R\right)
 = {\hat c_{0}}^2
 +{\hat c_{3}}^2
 \no
 &\det R_a
 =c_{a0}^2
 -c_{a3}^2,
 &\qquad &
 \frac{1}{2}\tr\left( R^T_a R_a\right)
 = c_{a0}^2
 +c_{a3}^2.
\end{alignat}
For $\hat R(0)$ and $R_a(0)$ to be real matrices, they need to satisfy
\begin{align}
 \tr\left(\hat R^T\hat R\right) \pm 2\det \hat R \ge 0
 , \qquad
 \tr\left( R^T_a R_a\right) \pm 2\det R_a \ge 0
 .
 \label{matrix-inequality}
\end{align}
Note also that $\tr(\hat R^T\hat R)\ge 0$ and $\tr(R^T_aR_a)\ge 0$.

Let us now discuss the inequalities for the BPS solution \eqref{BPSsolR}
in the case where all $\det R_a$ are non-zero.
The inequality \eqref{matrix-inequality} immediately
leads us to 
\begin{align}
 &\left(\frac{A\alpha_4}{\beta_4}\pm 1\right)\frac{1-A}{A+2}\det R_4 \ge 0,\\
 &\left(\frac{A\alpha_4}{\beta_4}\pm 1\right)\det R_4 \ge 0,\qquad
 \left(\frac{\alpha_6}{\beta_6}\pm 1\right)\det R_6 \ge 0,\qquad
 \left(\frac{\alpha_8}{\beta_8}\pm 1\right)\det R_8 \ge 0.
 \label{inequalities_468}
\end{align}
Therefore, $A$ needs to be in the range $-2<A\le 1$.
One can then obtain from $\frac{\alpha_6}{\beta_6}\det R_6>0$ and $\frac{\alpha_8}{\beta_8}\det R_8>0$
that
\begin{align}
 \det R_4\det R_6\det R_8 >0.
\end{align}
Then, this inequality and $\frac{A\alpha_4}{\beta_4}\det R_4>0$ 
together tell us that $A>0$,
and we arrive at \eqref{BPS-param-inequality}
by rewriting \eqref{inequalities_468}:
\begin{align*}
 0<A\le 1,
 \quad
 r_{\text{M5}}^4 A \ge |\det R_6 \det R_8|,
 \quad
 r_{\text{M5}}^4\frac{A+2}{3} \ge |\det R_4 \det R_8|,
 \quad
 r_{\text{M5}}^4\frac{A+2}{3} \ge |\det R_4 \det R_6|.
\end{align*}

These inequalities restrict the values of the angular momentum \eqref{BPS-angular-momentum}.
Since we have
\begin{align}
 A\left| \frac{\alpha_4}{\beta_4} \right|\ge 1,
 \qquad
 \left| \frac{\alpha_6}{\beta_6} \right|\ge 1,
 \qquad
 \left| \frac{\alpha_8}{\beta_8} \right|\ge 1
\end{align}
from \eqref{inequalities_468},
the angular momentum
satisfies the following inequalities:
\begin{align}
\frac{\eta_1\mu}{3}M_{12}&=-\frac{p^+}{3\cdot 18}\frac{\mu^2}{r^4_{\text{M5}}}\det R_4 \det R_6 \det R_8\frac{1-A}{A+2}\left( 2A\frac{\alpha^2_4}{\beta^2_4}+1 \right)\le 0,\no
\frac{\eta_4\mu}{6}M_{45}&=-\frac{p^+}{6\cdot 18}\frac{\mu^2}{r^4_{\text{M5}}}\det R_4 \det R_6 \det R_8 \left( A\frac{\alpha_4^2}{\beta_4^2}-1 \right)\le 0,\no
\frac{\eta_6\mu}{6}M_{67}&=-\frac{p^+}{6\cdot 18}\frac{\mu^2}{r^4_{\text{M5}}}\det R_4 \det R_6 \det R_8\frac{3}{A+2}\left( \frac{\alpha_6^2}{\beta_6^2}-1 \right)\le 0,\no
\frac{\eta_8\mu}{6}M_{89}&=-\frac{p^+}{6\cdot 18}\frac{\mu^2}{r^4_{\text{M5}}}\det R_4 \det R_6 \det R_8\frac{3}{A+2}\left( \frac{\alpha^2_8}{\beta^2_8}-1 \right)\le 0.
\end{align}
If $A<1$, then
$\frac{\eta_1\mu}{3}M_{12}$ and $\frac{\eta_1\mu}{6}M_{45}$
are negative definite.

\bibliographystyle{JHEP}
\bibliography{M5BPS}

\end{document}